\documentclass[conference]{IEEEtran}
\IEEEoverridecommandlockouts

\usepackage{times,amsmath,amssymb,amsfonts,amsthm,graphicx}
\usepackage{cite,color}
\usepackage{bm}
\usepackage{balance}
\usepackage{url}
\usepackage{subfigure}
\usepackage{booktabs}

\makeatletter
\newcommand*{\rom}[1]{\expandafter\@slowromancap\romannumeral #1@}
\makeatother

\def\alphabf{\boldsymbol \alpha}
\def\betabf{\boldsymbol \beta }

\def\deltabf{\boldsymbol \delta }

\def\thetabf{\boldsymbol \theta}

\def\Phibf{\boldsymbol \Phi}

\def\Lambdabf{\boldsymbol \Lambda}

\def\abf{{\bf a}}

\def\cbf{{\bf c}}

\def\ebf{{\bf e}}
\def\fbf{{\bf f}}
\def\gbf{{\bf g}}
\def\hbf{{\bf h}}

\def\nbf{{\bf n}}

\def\pbf{{\bf p}}

\def\sbf{{\bf s}}

\def\ubf{{\bf u}}
\def\vbf{{\bf v}}
\def\wbf{{\bf w}}
\def\xbf{{\bf x}}

\def\zbf{{\bf z}}

\def\xbf{{\bf x}}

\def\Abf{{\bf A}}
\def\Bbf{{\bf B}}

\def\Ibf{{\bf I}}

\def\Pbf{{\bf P}}

\def\Ac{{\cal A}}
\def\Bc{{\cal B}}

\def\Kc{{\cal K}}

\def\Nc{{\cal N}}

\def\Pc{{\cal P}}

\def\Sc{{\cal S}}
\def\Tc{{\cal T}}

\def\Xc{{\cal X}}
\def\Yc{{\cal Y}}

\def\ie{{\it i.e.,\ \/}}
\def\nn{\nonumber}

\def\Re{\mathfrak{R}\mathfrak{e}}
\def\Im{\mathfrak{I}\mathfrak{m}}

\def\mae{{\mathbb{E}}}

\newcommand{\olsi}[1]{\,\overline{\!{#1}}}

\theoremstyle{definition}

\newtheorem{lemma}{Lemma}

\newtheorem{assumption}{Assumption}
\newtheorem{proposition}{Proposition}

\newenvironment{mylist}%
{\begin{list}{}%
    {%
      \setlength{\itemindent}{-5pt}%
      \setlength{\leftmargin}{12pt}%
      \setlength{\parsep}{\parskip}
      \setlength{\labelsep}{5pt}
      \setlength{\itemsep}{2pt}}}%
  {\end{list}}

\begin{document}

\title{SegOTA: Accelerating Over-the-Air Federated Learning with Segmented Transmission}

\author{ Chong Zhang$^{\star}$, Min Dong$^{\dagger}$, Ben Liang$^{\star}$, Ali Afana$^{\ddagger}$, Yahia Ahmed$^{\ddagger}$\\
\normalsize $^{\star}$Department of Electrical and Computer Engineering, University of Toronto, Canada   \\
$^{\dagger}$Department of Electrical, Computer and Software Engineering, Ontario Tech University, Canada,
$^{\ddagger}$Ericsson Canada\thanks{This work was supported in part by Ericsson, the Natural Sciences and Engineering Research Council of Canada, and Mitacs.}
}%

\maketitle

\begin{abstract}
Federated learning (FL) with over-the-air computation
efficiently utilizes the communication resources, but it can still experience significant latency when each device transmits a large number of model parameters to the server. This paper proposes the Segmented Over-The-Air (SegOTA) method for FL, which reduces  latency by partitioning devices into groups and letting each group transmit only one segment of the model parameters in each communication round. Considering a multi-antenna server, we model the SegOTA transmission and reception process to establish an upper
bound on the expected model learning optimality gap. We minimize this upper bound, by formulating the per-round online optimization  of device grouping and joint transmit-receive beamforming, for which we derive efficient closed-form solutions. Simulation results show that our proposed SegOTA substantially outperforms the conventional full-model OTA approach and other common alternatives.
\end{abstract}

\section{Introduction}
\label{sec:intro}

\begin{figure*}[!htbp]
\centering
\hspace*{1em}\includegraphics[scale=.28]{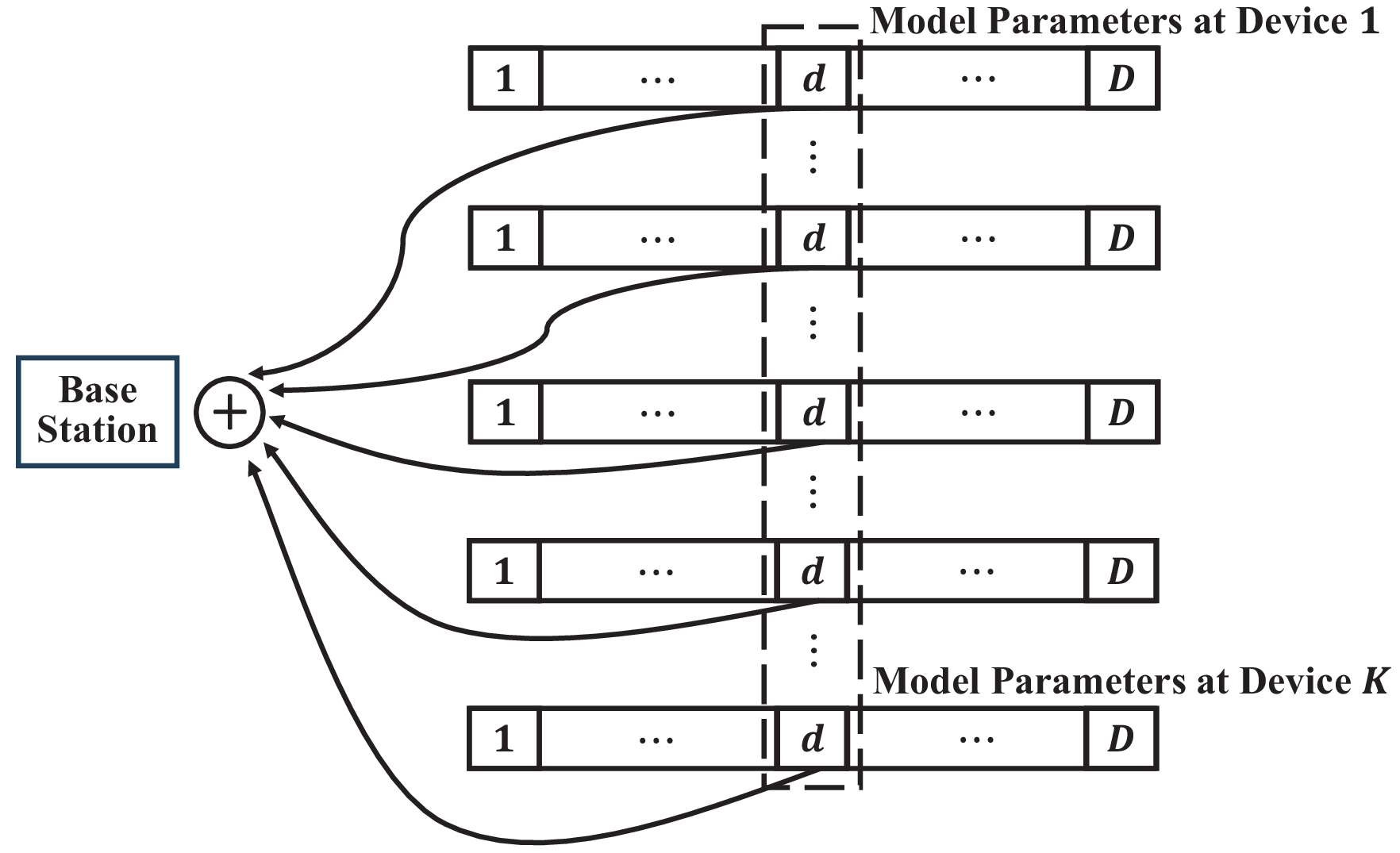}\hspace*{.45em}\includegraphics[scale=.28]{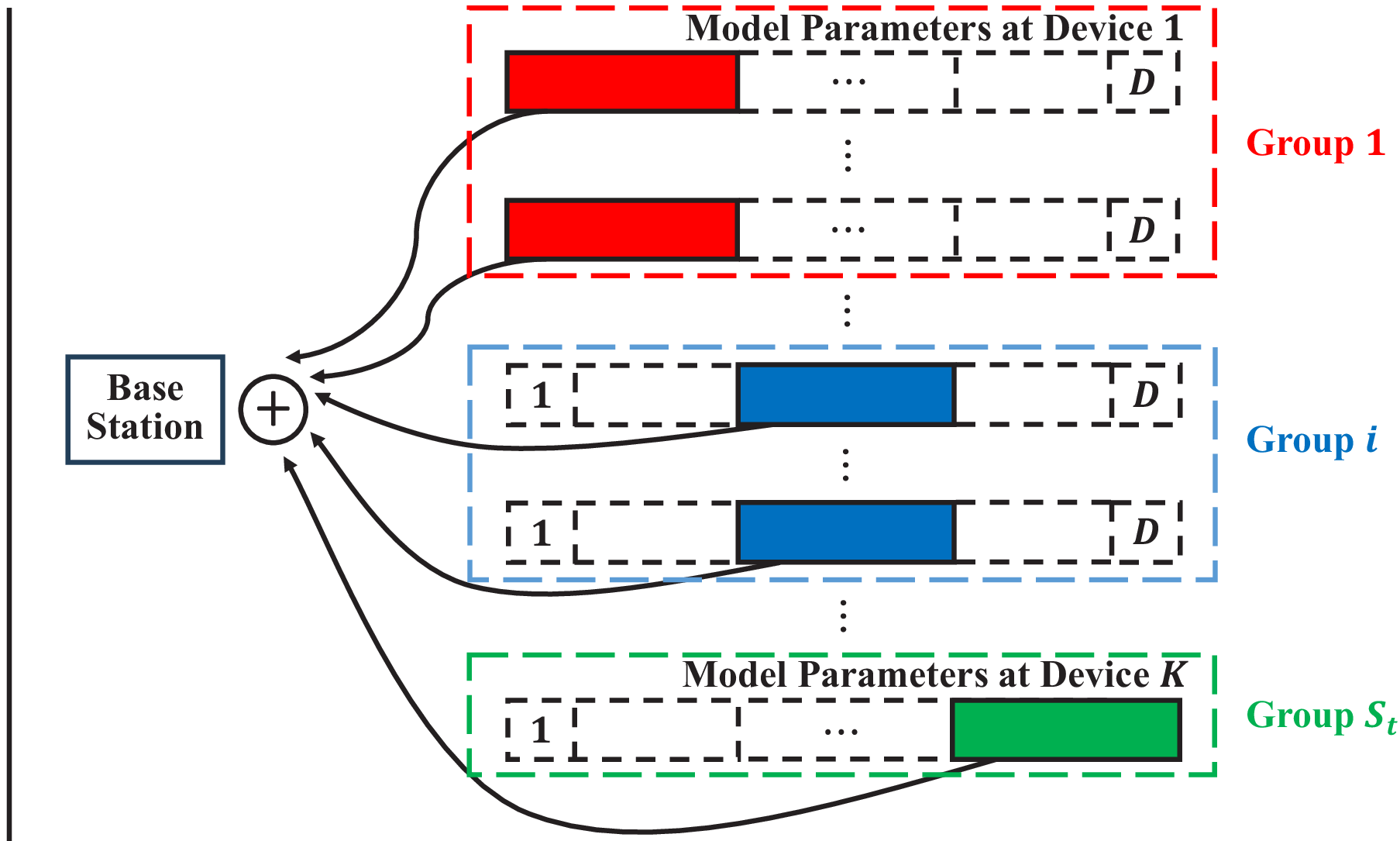}
\caption{Uplink analog OTA aggregation for wireless FL. Left: traditional full-model OTA approach; Right: proposed SegOTA
(Each colored box represents a model segment that consists of $I_t$ parameters; different colors indicate different segments).}
\label{fig1:full_split_model_transmission}\vspace*{-1em}
\end{figure*}

Federated learning (FL) \cite{Mcmahan&etal:2017} enables multiple worker devices to collaboratively train a machine learning model using
their local datasets, with a parameter server (PS) aggregating their local updates into a global model.
In wireless FL, the PS often is hosted by a base station (BS) \cite{Zhu&etal:2020}.
However, limited wireless resources and signal distortion in wireless links degrade the performance of wireless FL,
making efficient communication design a necessity.

Most prior wireless FL works focused on improving the communication efficiency in uplink aggregation of local model parameters from the devices to the BS
\cite{Du&etal:2020TSP,Amiri&etal:TWC2021,Wang&etal:JSAC2022,
Zhu&etal:TWC2020,Zhang&Tao:TWC2021,Wang&etal:ToN2024,
Yang&etal:TWC2020,Liu&etal:TWC2021,Kim&etal:TWC2023,Kalarde&etal:2024,Zhang&etal:SPAWC24}.
Early works \cite{Du&etal:2020TSP, Amiri&etal:TWC2021,Wang&etal:JSAC2022}
studied digital transmission-then-aggregation schemes using orthogonal channels,
which can consume large bandwidth and cause high latency with many devices.
Later,  analog transmission-and-aggregation schemes were proposed
\cite{Zhu&etal:TWC2020,Zhang&Tao:TWC2021,Wang&etal:ToN2024}, which adopt analog modulation and superposition for
over-the-air computation of local parameters via the multiple
access channel.
Analog schemes result in significant communication savings and lower latency compared with digital approaches. However, these analog schemes are designed for single-antenna BSs, making their solutions and convergence analysis unsuitable for the multi-antenna BSs commonly used in practical wireless systems.

In the multi-antenna communication, beamforming plays a critical role in improving communication quality in wireless networks.
Receive beamforming was considered in \cite{Chen&etal:IoT2018,Zhu&Huang:Globecom18} to boost performance of the uplink analog
over-the-air computation.
Various uplink beamforming designs have since been proposed to
improve the training performance of wireless FL \cite{Yang&etal:TWC2020,Liu&etal:TWC2021,Kim&etal:TWC2023,Kalarde&etal:2024,Zhang&etal:SPAWC24}.
These works demonstrate that well-designed beamforming schemes can significantly enhance the over-the-air computation
for wireless FL.

The existing uplink analog over-the-air works \cite{Zhu&etal:TWC2020,Zhang&Tao:TWC2021,Wang&etal:ToN2024,Yang&etal:TWC2020,Liu&etal:TWC2021,Kim&etal:TWC2023,Kalarde&etal:2024,Zhang&etal:SPAWC24} typically adopt the traditional full-model transmission approach illustrated in Fig.~\ref{fig1:full_split_model_transmission}(left).
In each channel use, all devices simultaneously send parameters from the same location in their model parameter vectors to the BS, which aggregates these parameters via over-the-air computation. However, this full-model transmission approach can lead to substantial latency when the model parameter vector is long, which degrades the overall performance of wireless FL.

To address this issue, we propose the Segmented Over-The-Air (SegOTA) method for wireless FL,
as illustrated in Fig.~\ref{fig1:full_split_model_transmission}(right). SegOTA divides the locations of
a model parameter vector into equal-sized segments and assigns the transmission task of each segment
to a group of devices.
By allowing simultaneous transmission of parameters in different segments, SegOTA can substantially reduce the communication latency, while maintaining satisfactory learning performance by allowing the BS to aggregate and update global parameters for each segment. However, it also introduces wireless interference among the different segments, which requires careful transmission design to balance the tradeoff between communication efficiency and OTA computation accuracy, in order to optimize the overall FL performance.

The main contribution of this paper is summarized as follows:
\begin{mylist}
\item We propose a novel SegOTA method to allow simultaneous transmission of parameters from multiple segments in wireless FL. We formulate an
optimization problem to carefully manage the inter-segment interference through device grouping, BS receive beamforming, and device transmit power control, in order to minimize the expected model optimality gap after a given number of FL communication rounds. As far as we are aware, this is the first study on segmented OTA transmission in FL.

\item We analyze the SegOTA transmission
and reception processes and derive an upper bound
on the expected model optimality gap. We show that minimization of this
bound can be decomposed into per-round online optimization
problems of device grouping and uplink joint transmit-receive
beamforming. We apply a spherical {\it k}-means method
for device grouping and then optimize joint transmit-receive
beamforming, obtaining efficient closed-form solutions at each
round. The proposed solution is guaranteed to converge to a stationary point.

\item Simulation under typical wireless network settings
shows that SegOTA substantially outperforms the conventional
approach of full-model OTA aggregation, as well as other alternatives such as segmented OTA with the popular zero-forcing beamforming.
\end{mylist}

\allowdisplaybreaks
\section{Wireless FL System Model}\label{sec:system_prob}

We consider an FL system consisting of a server and
$K$ worker devices that collaboratively train a machine learning model.  Let  $\Kc_{\text{tot}} = \{1, \ldots, K\}$ denote the total set of devices.
Each device $k \in \Kc_{\text{tot}}$ holds a local training dataset of size $A_k$, denoted by
$\Ac_{k} = \{(\abf_{k,i},y_{k,i}): 1 \le i \le A_k\}$, where
$\abf_{k,i}\in\mathbb{R}^{b} $ is the $i$-th data feature vector and $y_{k,i}$ is the label for this data sample.
Let $\thetabf\in\mathbb{R}^{D}$ be the parameter vector of the
machine learning model, which has $D$ parameters.
The local training loss function that represents the training error
at device $k$ is given by
\begin{align}
F_{k}(\thetabf) = \frac{1}{A_k}\sum_{i=1}^{A_k} L(\thetabf;\abf_{k,i},y_{k,i})  \nn
\end{align}
where $L(\cdot)$ is the sample-wise training loss corresponding to each data sample.
The global training loss function is the weighted sum of the local loss function $F_{k}(\thetabf)$ of each device $k$,
expressed as
\begin{align}
F(\thetabf) = \frac{1}{\sum_{k=1}^{K}A_{k}}\sum_{k=1}^{K}A_kF_{k}(\thetabf).  \label{eq_global_loss_original}
\end{align}
The learning objective is to find the optimal global model $\thetabf^\star$ that minimizes $F(\thetabf)$.

The $K$ devices communicate with the server via separate downlink and uplink channels to exchange the model training information iteratively.
The FL training procedure in each communication round  $t=0,1,\ldots$ is given as follows:
\begin{mylist}
\item \emph{Downlink broadcast}: The server sends the parameter vector of the current global model $\thetabf_t$
    to all $K$ devices. We make the common assumption that the downlink channel is error-free.
\item  \emph{Local model update}: Device $k$ divides $\Ac_k$ into smaller mini-batches, and applies the standard mini-batch stochastic gradient descent (SGD) algorithm with $J$ iterations to generate an updated local model
    based on $\thetabf_{t}$.
    Let $\thetabf^{\tau}_{k,t} $ be the local model update by device $k$ at iteration $\tau \in \{0,\ldots,J-1\}$, with
    $\thetabf^{0}_{k,t} = \thetabf_{t}$, and let  $\Bc^{\tau}_{k,t}$ denote the mini-batch at iteration  $\tau$.
    Then, the local model update is given by
    \begin{align}
    \thetabf^{\tau+1}_{k,t} & = \thetabf^{\tau}_{k,t} - \eta_t \nabla F_{k}(\thetabf^{\tau}_{k,t}; \Bc^{\tau}_{k,t}) \nn\\
    & = \thetabf^{\tau}_{k,t} - \frac{\eta_t}{|\Bc^{\tau}_{k,t}|}\sum_{(\abf,y)\in\Bc^{\tau}_{k,t}}\!\!\!\nabla L(\thetabf^{\tau}_{k,t}; \abf,y)
    \label{eq_sgd}
    \end{align}
    where $\eta_{t}$ is the  learning rate in communication round $t$,
    and $\nabla F_{k}$ and $\nabla L$ are the gradient functions  with respect to (w.r.t.) $\thetabf^{\tau}_{k,t}$.
    After  $J$ iterations, device $k$ obtains the updated local model $\thetabf^{J}_{k,t}$.

\item \emph{Uplink aggregation}: The devices send their updated local models $\{\thetabf^{J}_{k,t}\}_{k\in\Kc_{\text{tot}}}$ to the server
     through the uplink channels. The server aggregates $\thetabf^{J}_{k,t}$'s to generate an updated global model $\thetabf_{t+1}$ for the next communication round $t+1$.
In the existing full-model OTA approach \cite{Zhu&etal:TWC2020,Zhang&Tao:TWC2021,Wang&etal:ToN2024,Yang&etal:TWC2020,Liu&etal:TWC2021,Kim&etal:TWC2023,Kalarde&etal:2024,Zhang&etal:SPAWC24}, this consists of analog transmission of each of the parameters in the $\thetabf^{J}_{k,t}$ vector by device $k$, and OTA aggregation by the BS, as shown in Fig.~1(left). In this paper, we will propose a segmented OTA approach to improve the communication efficiency of this step.

\end{mylist}

For the model exchange between the server and devices through a wireless system, we assume the server is hosted by a BS equipped with
$N$ antennas, and each device has a single antenna.
In this paper, we propose a segmented transmission approach for uplink OTA aggregation and optimize the corresponding uplink beamforming to accelerate the FL training convergence.

\section{Uplink Segmented Over-the-Air Aggregation}
\label{sec:SegOTA}

We propose an efficient uplink aggregation approach, named SegOTA. Under SegOTA, each device only sends one segment of its $D$ local parameters to the BS in each communication round, shown in
Fig.~\ref{fig1:full_split_model_transmission}(right)  as a colored segment.

In particular, at the beginning of communication round $t$,
the BS partitions the $K$ devices into $S_t$ groups,
which remain unchanged during this round.
Let $\Kc_{i,t}$ denote the set of devices of group
$i\in\{1,\ldots,S_t\}$ in round $t$, with $K_{i,t}\triangleq |\Kc_{i,t}|$.
The BS also divides the model parameter vector into $S_t$ equal-sized segments,
with each segment having a length of
$I_t \triangleq \lceil\frac{D}{S_t}\rceil$.
If $D$ is not a multiple of $S_t$,
the last segment will be padded with zero.
Let $\Sc_{t} \triangleq \{1, \ldots, S_{t}\}$ be the index set of model segments in round $t$.
Devices in group $\Kc_{i,t}$ are assigned to send segment $\hat{m}(i, t)\in\Sc_{t}$ to the BS. Each group is assigned a unique segment, which can be either at random
or in a round-robin fashion.
The BS then aggregates the received local segments from the devices in group $\Kc_{i,t}$ to update segment $\hat{m}(i, t)$ of the global
model $\thetabf_{t+1}$ for the next communication round $t+1$.
Below, we detail the formulation of uplink aggregation under the proposed SegOTA.

Let $\sbf^{k,J}_{m,t}\in\mathbb{R}^{I_t}$ denote the segment $m$ of the local model update $\thetabf^{J}_{k,t}$ at device $k$
in communication round $t$.
For efficient transmission,  we represent $\sbf^{k,J}_{m,t}\in\mathbb{R}^{I_t}$ using an equivalent complex vector $\tilde{\sbf}^{k,J}_{m,t}$,
whose real and imaginary parts contain the first and second halves of the elements in $\sbf^{k,J}_{m,t}$, respectively.
That is, $\tilde{\sbf}^{k,J}_{m,t} = \tilde{\sbf}^{k,J\text{re}}_{m,t} + j\tilde{\sbf}^{k,J\text{im}}_{m,t}\in\mathbb{C}^{\frac{I_t}{2}}$, where $\tilde{\sbf}^{k,J\text{re}}_{m,t}$ contains the first
$\frac{I_t}{2}$ elements in $\sbf^{k,J}_{m,t}$ and $\tilde{\sbf}^{k,J\text{im}}_{m,t}$ contains the other $\frac{I_t}{2}$ elements.

We denote the channel from device $k$ to the BS in communication round $t$ by $\hbf_{k,t}$
and assume it is known at the BS.
We denote the transmit beamforming weight at device $k$ in round $t$
by $a_{k,t}\in\mathbb{C}$.
Device $k$ in group $i$ applies $a_{k,t}$ to the normalized complex model segment
$\frac{\tilde{\sbf}^{k,J}_{\hat{m}(i, t),t}}{\|\tilde{\sbf}^{k,J}_{\hat{m}(i, t),t}\|}$,
and all $K$ devices send their respective segments simultaneously
to the BS with $\frac{I_t}{2}$ channel uses.
Let $\frac{\tilde{s}^{k,J}_{\hat{m}(i, t)l,t}}{\|\tilde{\sbf}^{k,J}_{\hat{m}(i, t),t}\|}$ be the $l$-th element in segment $\frac{\tilde{\sbf}^{k,J}_{\hat{m}(i, t),t}}{\|\tilde{\sbf}^{k,J}_{\hat{m}(i, t),t}\|}$
sent in the $l$-th channel use.
The received signal vector at the BS in the $l$-th channel use
is given by
\begin{align}
\vbf_{l,t} = \sum_{i=1}^{S_t}\sum_{k\in\Kc_{i,t}}\hbf_{k,t}a_{k,t}
\frac{\tilde{s}^{k,J}_{\hat{m}(i, t)l,t}}{\|\tilde{\sbf}^{k,J}_{\hat{m}(i, t),t}\|} + \ubf_{l,t} \nn
\end{align}
where $\ubf_{l,t}\sim \mathcal{CN}({\bf 0}, \sigma^2\Ibf)$  is the receiver additive white Gaussian noise vector with variance $\sigma^2$.

The BS applies receive beamforming on $\vbf_{l,t}$ to aggregate the local segments $\{\tilde{\sbf}^{k,J}_{\hat{m}(i, t),t}\}_{k\in\Kc_{i,t}}$ from each group $i$.
Let $\wbf_{i,t}\in\mathbb{C}^{N}$ denote the receive beamforming vector at the BS for group $i$ in communication round $t$,
which is normalized as $\|\wbf_{i,t}\|^2 = 1$.
The effective channel from device $k\in\Kc_{i,t}$ to the BS after applying $\wbf_{i,t}$
is given by $\alpha_{k,t} \triangleq \frac{\wbf^\textsf{H}_{i,t}\hbf_{k,t}a_{k,t}}{\|\tilde{\sbf}^{k,J}_{\hat{m}(i, t),t}\|}$.
Then, the post-processed received signal vector for
the aggregated local segments
$\{\tilde{\sbf}^{k,J}_{\hat{m}(i, t),t}\}_{k\in\Kc_{i,t}}$ over the $\frac{I_t}{2}$ channel uses is
\begin{align}
\zbf_{\hat{m}(i, t),t}\! = & \!\!\sum_{k\in\Kc_{i,t}}\!\!\!\alpha_{k,t}\tilde{\sbf}^{k,J}_{\hat{m}(i, t),t}\! +\! \sum_{j\neq i}\!\sum_{q\in\Kc_{j,t}}\!\!\wbf_{i,t}^\textsf{H}\hbf_{q,t}a_{q,t}\frac{\tilde{\sbf}^{q,J}_{\hat{m}(j, t),t}}{\|\tilde{\sbf}^{q,J}_{\hat{m}(j, t),t}\|} \nn\\
&\qquad  + \nbf_{\hat{m}(i, t),t}
\label{eq_zmt_post_process}
\end{align}
where $\nbf_{\hat{m}(i, t),t}\in\mathbb{C}^{\frac{I_t}{2}}$ is the post-processed receiver noise vector with  the $l$-th element being $\wbf^\textsf{H}_{i,t}
\ubf_{l,t}$, and the last term is the inter-segment interference.

After receive beamforming, the BS receiver finally performs scaling to obtain the global model update.
Let $\sbf_{m,t}\in\mathbb{R}^{I_t}$ denote segment $m$ of the global model update $\thetabf_{t}$
in communication round $t$, and let $\tilde{\sbf}_{m,t}$ be the equivalent complex representation of $\sbf_{m,t}$.
Let $\hat{i}(m,t)$ represent the device group that transmits segment $m$ in round $t$.
The BS scales $\zbf_{m,t}$ by $\alpha^{\text{s}}_{m,t} \triangleq \sum_{k\in\Kc_{\hat{i}(m,t),t}}\alpha_{k,t}$
 to obtain
segment $m$ of the global model update $\thetabf_{t+1}$ for the next round $t+1$:
\begin{align}
\tilde{\sbf}_{m,t+1} = \frac{\zbf_{m,t}}{\alpha^{\text{s}}_{m,t}}.
\label{eq_scale_seg}
\end{align}
Based on \eqref{eq_zmt_post_process}\eqref{eq_scale_seg},
we obtain the following updating equation for segment $\tilde{\sbf}_{m,t}$
in each round:
\begin{align}
& \tilde{\sbf}_{m,t+1} = \tilde{\sbf}_{m,t} + \sum_{k\in\Kc_{\hat{i}(m,t),t}}\rho_{k,t}\Delta\tilde{\sbf}^{k}_{m,t}+ \tilde{\nbf}_{m,t} \nn\\
&\qquad + \frac{1}{\alpha^{\text{s}}_{m,t}}\!\sum_{j\neq \hat{i}(m,t)}\sum_{q\in\Kc_{j,t}}\wbf_{\hat{i}(m,t),t}^\textsf{H}
\hbf_{q,t}a_{q,t}  \frac{\tilde{\sbf}^{q,J}_{\hat{m}(j, t),t}}{\|\tilde{\sbf}^{q,J}_{\hat{m}(j, t),t}\|}
\label{eq_global_update_0}
\end{align}
where $\Delta\tilde{\sbf}^{k}_{m,t} \triangleq \tilde{\sbf}^{k,J}_{m,t}
- \tilde{\sbf}^{k,0}_{m,t}$ is the difference in the local segment update after the local training,
$\rho_{k,t}  \triangleq \frac{\alpha_{k,t}}{\alpha^{\text{s}}_{m,t}}$
represents the weight of each device $k$ in group $\hat{i}(m,t)$,
and $\tilde{\nbf}_{m,t}  \triangleq  \frac{\nbf_{m,t}}{\alpha^{\text{s}}_{m,t}}$ is
post-processed receiver noise vector.

Finally, segment $m$ of the global model update $\thetabf_{t+1}$,
\ie $\sbf_{m,t+1}$, can be recovered from its complex version as
$\sbf_{m,t+1}\!=[\Re{\{\tilde{\sbf}_{m,t+1}\}^\textsf{T}}\!, \Im{\{\tilde{\sbf}_{m,t+1}\}^\textsf{T}}]^\textsf{T}$.

\section{SegOTA Design Optimization}
\subsection{Joint Optimization Formulation}\label{sec:joint_problem} \label{sec:upper_bound}
We focus on the uplink communication design involved in SegOTA, aiming  to maximize the FL training convergence rate.
For effective SegOTA, we
 consider joint transmit-receive beamforming in the uplink, where the device transmit beamforming weights $\{a_{k,t}\}_{k\in\Kc_{i,t}}$ and the BS receive beamforming vector $\wbf_{i,t}$ are designed jointly for each group $i$ in communication round $t$.
In communication round $t$, the model segments $\{\tilde{\sbf}^{k,J}_{\hat{m}(i, t),t}\}_{k\in\Kc_{i,t}}$ from devices within the same group $i$ need to be combined coherently, as indicated in the first term of \eqref{eq_zmt_post_process}.
To achieve this, we phase-align the effective channels $\{a_{k,t}\}_{k\in\Kc_{i,t}}$ among  the devices in group $i$ via the joint transmit-receive beamforming.
In particular, we set the transmit beamforming weight at device $k$ in group $i$ as $a_{k,t} =\sqrt{p_{k,t}}  \frac{\hbf_{k,t}^\textsf{H}\wbf_{i,t}}
{|\hbf^\textsf{H}_{k,t}\wbf_{i,t}|}$,
where $p_{k,t}$ is the transmit power used for sending the entire segment at device $k$.
Then, the effective channel of this device $k$ is given by
\begin{align}
\alpha_{k,t} = \frac{\wbf_{i,t}^\textsf{H}\hbf_{k,t}a_{k,t}}{\|\tilde{\sbf}^{k,J}_{\hat{m}(i, t),t}\|} = \frac{\sqrt{p_{k,t}}|\hbf^\textsf{H}_{k,t}\wbf_{i,t}|}{\|\tilde{\sbf}^{k,J}_{\hat{m}(i, t),t}\|}, \ k \in \Kc_{i,t}.  \label{eq_effect_channel}
\end{align}

Given a total number of device groups $S_t$,
our goal is to optimize the device grouping and joint transmit-receive beamforming to
minimize the expected optimality gap $\mae[\|\thetabf_{T}- \thetabf^\star\|^2]$ after $T$ communication rounds.
Let $\pbf_t\triangleq[\pbf^\textsf{T}_{1,t}, \ldots, \pbf^\textsf{T}_{S_t,t}]^\textsf{T}$, where $\pbf_{i,t}\in\mathbb{R}^{K_{i,t}}$ contains the transmit power of each device in group $i$ of round $t$, $p_{k,t}$, $k\in\Kc_{i,t}$. Also, $\wbf_t\triangleq[\wbf_{1,t}^\textsf{H}, \ldots, \wbf_{S_t,t}^\textsf{H}]^\textsf{H}\in\mathbb{C}^{S_tN}$
contains the BS receive beamforming vectors for all $S_t$ groups in round $t$.
Let $\Tc\triangleq \{0,\ldots,T-1\}$.
Then, the joint optimization problem is formulated as
\begin{align}
\Pc_{o}: \,\, & \min_{\{ \{\Kc_{i,t}\}^{S_t}_{i=1}, \wbf_t, \pbf_t\}^{T-1}_{t=0}} \,\, \mae[\|\thetabf_{T}- \thetabf^\star\|^2] \nn\\
\text{s.t.} \ \ & \Kc_{i,t}\bigcap\Kc_{j,t} = \varnothing,\quad i\neq j,\,\, i,j\in\Sc_t,\,\, t\in\Tc,   \label{constra_grouping_1}\\
& \bigcup\limits_{i\in\Sc_t}\Kc_{i,t} = \Kc_{\text{tot}}, \quad t\in\Tc,\label{constra_grouping_2}\\
& p_{k,t} \le I_tP_k, \quad k\in \Kc_{\text{tot}},\,\, t\in\Tc,\label{constra_device_power}\\
& \|\wbf_{i,t}\|^2 = 1, \quad i\in \Sc_t,\,\, t\in\Tc \label{constra_receive_BF}
\end{align}
where $\mae[\cdot]$ is  taken w.r.t.\ receiver noise and
mini-batch local data samples  at each device,
\eqref{constra_grouping_1} and \eqref{constra_grouping_2} are the device grouping constraints,
$P_k$ is the average transmit power budget at device $k$ for sending a signal
and constraint \eqref{constra_device_power} specifies the per-device average transmit power budget for sending the entire model segment
in each communication round.

Problem $\Pc_{o}$ is a $T$-horizon stochastic optimization problem. To tackle this challenging problem, we first analyze the training convergence rate and develop a more tractable upper bound on $\mae[\|\thetabf_{T}- \thetabf^\star\|^2]$.
Then, we propose an efficient scheme for device grouping and joint transmit-receive beamforming to minimize this upper bound.

\subsection{SegOTA Training Convergence Analysis}\label{sec:SegOTA_convrg_analysis}
To analyze the FL training convergence speed, we make the following assumptions. They are commonly used in the existing literature for conventional full-model OTA aggregation, and here we have extended them to segmented OTA.
\begin{assumption}\label{assump_smooth}
The local loss function $F_{k}(\cdot)$ is $L$-smooth and $\lambda$-strongly convex,
$k\in\Kc_{\text{tot}}$.
\end{assumption}

Moreover, let $\nabla F$ denote the gradient of the global loss function given in
\eqref{eq_global_loss_original} w.r.t. $\thetabf_{t}$.
Denote segments $m$ of gradients $\nabla F(\thetabf_{t})$ and $\nabla F_{k}(\thetabf^{\tau}_{k,t})$
respectively by $\nabla F^{m}(\thetabf_{t})$ and $\nabla F^m_{k}(\thetabf^{\tau}_{k,t})$.
We then make Assumption~\ref{assump_bound_diff} below.

\begin{assumption}\label{assump_bound_diff}
Bounded gradient divergence: for all $t$, $\tau$, and $m$,
$\mae[\| \nabla F^{m}(\thetabf_{t}) - \sum_{k}c_{k}
\nabla F^m_{k}(\thetabf^{\tau}_{k,t})  \|^2] \leq \phi$,
for some $\phi \ge 0$ and $0\leq c_k \leq 1$ such that
$\sum_{k}c_{k}=1$.
Furthermore, for all $t$, $\tau$, $k$, and $m$,
 $\mae[\| \nabla F_{k}^{m}(\thetabf^{\tau}_{k,t}) -
\nabla F_{k}^{m}(\thetabf^{\tau}_{k,t}; \Bc^{\tau}_{k,t})\|^2]
\leq \mu$, for some $\mu \ge 0$.
\end{assumption}

We apply $a_{k,t}$ and $\alpha_{k,t}$ given at the beginning of Section~\ref{sec:upper_bound}
into the segment updating equation \eqref{eq_global_update_0}
and further define $\Delta\tilde{\sbf}_{m,t} \triangleq \sum_{k\in\Kc_{\hat{i}(m,t),t}}\rho_{k,t}\Delta\tilde{\sbf}^{k}_{m,t}$.
Moreover, let $\tilde{\ebf}_{m,t}$ be the fourth term for the inter-segment interference
on the right-hand side of \eqref{eq_global_update_0}.
Stacking all $S_t$ segments, $\tilde{\sbf}_{m,t+1}$,
$m\in\Sc_t$, together, following \eqref{eq_global_update_0}, we express the entire global model update $\tilde{\thetabf}_{t+1}$ from $\tilde{\thetabf}_{t}$ as
\begin{align}
\tilde{\thetabf}_{t+1} = \tilde{\thetabf}_{t} + \Delta\tilde{\thetabf}_{t} + \tilde{\nbf}_{t}
+ \tilde{\ebf}_{t}
\label{eq_entire_model_update}
\end{align}
where $\Delta\tilde{\thetabf}_{t}$ is the vector that stacks
$\Delta\tilde{\sbf}_{m,t}$, $m\in\Sc_t$,
and $\tilde{\nbf}_{t}$ and $\tilde{\ebf}_{t}$ are similarly defined.

We analyze the expected optimality gap, $\mae[\|\thetabf_{t+1}- \thetabf^\star\|^2]$ at round $t+1$.
Based on \eqref{eq_entire_model_update}, we can show that its upper bound is a function of $\mae[\|\thetabf_{t}- \thetabf^\star\|^2]$.
In particular,
we first consider an ideal centralized model training procedure using the full gradient descent training algorithm
and all the device datasets.
We assume the BS has the device datasets and implements this centralized training procedure without
exchanging any model updating information with the devices.
Let $\vbf^{\tau}_{t}$ be the model update at iteration $\tau \in \{0,\ldots,J-1\}$, with
$\vbf^{0}_{t} = \thetabf_{t}$.
The ideal centralized model update is given by
\begin{align}
\vbf^{\tau+1}_{t} = \vbf^{\tau}_{t} - \eta_{t}\nabla F(\vbf^{\tau}_{t}). \label{eq_ideal_GD}
\end{align}
Let $\tilde{\thetabf}^\star$, $\nabla \tilde{F}^m_{k}(\thetabf^{\tau}_{k,t})$,
$\tilde{\vbf}^{\tau}_{t}$, and $\nabla \tilde{F}(\vbf^{\tau}_{t})$ respectively denote the equivalent complex representations of $\thetabf^\star$,
$\nabla F^m_{k}(\thetabf^{\tau}_{k,t})$, $\vbf^{\tau}_{t}$, and $\nabla F(\vbf^{\tau}_{t})$.
Based on \eqref{eq_entire_model_update}\eqref{eq_ideal_GD}, we have
\begin{align}
&\!\!\!\! \tilde{\thetabf}_{t+1} - \tilde{\thetabf}^\star \nn\\
&\!\!\!\!\! =\! \tilde{\thetabf}_{t} - \eta_{t}\!\!\sum^{J-1}_{\tau=0}\!\nabla\! \tilde{F}(\vbf^{\tau}_{t}) - \tilde{\thetabf}^\star \! + \eta_{t}\!\!\sum^{J-1}_{\tau=0}\!\nabla\! \tilde{F}(\vbf^{\tau}_{t}) + \Delta\tilde{\thetabf}_{t} + \tilde{\nbf}_{t}
 + \tilde{\ebf}_{t}\nn\\
&\!\!\!\!\! =\! \tilde{\vbf}^{J}_{t} - \tilde{\thetabf}^\star + \eta_{t}\sum^{J-1}_{\tau=0}\nabla \tilde{F}(\vbf^{\tau}_{t}) + \Delta\tilde{\thetabf}_{t} + \tilde{\nbf}_{t} + \tilde{\ebf}_{t} \nn\\
&\!\!\!\!\! =\! \tilde{\vbf}^{J}_{t}\! -\! \tilde{\thetabf}^\star\! +\! \underbrace{\eta_{t}\!\sum^{J-1}_{\tau=0}\!\nabla \tilde{F}(\vbf^{\tau}_{t})
\!-\! \Delta\bar{\thetabf}_{t}}_{ \triangleq\tilde{\alphabf}_{t}}\! +\! \underbrace{\Delta\bar{\thetabf}_{t}\! +\! \Delta\tilde{\thetabf}_{t}}_{ \triangleq\tilde{\betabf}_{t}}\!   + \!\underbrace{\tilde{\nbf}_{t}\! +\! \tilde{\ebf}_{t}}_{ \triangleq\tilde{\deltabf}_{t}}
 \label{eq_frame_update_derive}
\end{align}
where $\Delta\bar{\thetabf}_{t}$ stacks
$\eta_t\sum_{k\in\Kc_{\hat{i}(m,t),t}}\rho_{k,t}\sum_{\tau=0}^{J-1} \nabla \tilde{F}_{k}^{m}(\thetabf^{\tau}_{k,t})$, $m\in\Sc_t$.
Following the above, we have
\begin{align}
&\mae[\|\tilde{\thetabf}_{t+1} - \tilde{\thetabf}^\star\|^2] = \mae[\|\tilde{\vbf}^{J}_{t} - \tilde{\thetabf}^\star + \tilde{\alphabf}_{t}+\tilde{\betabf}_{t}+\tilde{\deltabf}_{t}\|^2] \nn\\
& \leq \mae[(\|\tilde{\vbf}^{J}_{t} - \tilde{\thetabf}^\star\|+\|\tilde{\alphabf}_{t}\|+\|\tilde{\betabf}_{t}\|+\|\tilde{\deltabf}_{t}\|)^2]\nn\\
& \stackrel{(a)}{\leq} 4\big(\mae[\|\tilde{\vbf}^{J}_{t} - \tilde{\thetabf}^\star\|^2]
+ \mae[\|\tilde{\alphabf}_{t}\|^2] + \mae[\|\tilde{\betabf}_{t}\|^2] + \mae[\|\tilde{\deltabf}_{t}\|^2]\big)
\label{eq_bound0}
\end{align}
where $(a)$ is based on a specific case of the Cauchy–Schwarz inequality
$( \sum^{G}_{i=1}x_{i} )^2 \leq G\sum^{G}_{i=1}x^2_{i}, \forall x_{i}\in\mathbb{R}$, for some $G\in\mathbb{N}^{+}$.
We upper bound each term in \eqref{eq_bound0} below.

We first obtain an upper bound for $\mae[\|\tilde{\vbf}^{J}_{t} - \tilde{\thetabf}^\star\|^2] $ in Lemma~\ref{lemma1}.
The proof uses the same technique as in \cite[Lemma 2]{Bhuyan&etal:2023} and thus is omitted.
\begin{lemma}[Bounding ${\mae[\|\tilde{\vbf}^{J}_{t} - \tilde{\thetabf}^\star\|^2] }$]\label{lemma1}
 Consider SegOTA described in Section~\ref{sec:SegOTA} and the ideal centralized training described in
 Section~\ref{sec:SegOTA_convrg_analysis}.
 For $\eta_t<\frac{1}{L}$, $\forall t\in\Tc$,
 under Assumption~\ref{assump_smooth}, $\mae[\|\tilde{\vbf}^{J}_{t} - \tilde{\thetabf}^\star\|^2]$ is upper bounded as
\begin{align}
& \mae[\|\tilde{\vbf}^{J}_{t} - \tilde{\thetabf}^\star\|^2]
 \leq
(1-\eta_t\lambda)^{2J}\mae[\|\tilde{\thetabf}_{t} - \tilde{\thetabf}^\star\|^2], \ t\in\Tc. \label{eq_bound1}
\end{align}
\end{lemma}

Next, we bound the terms $\mae[\|\tilde{\alphabf}_{t}\|^2]$, $\mae[\|\tilde{\betabf}_{t}\|^2]$,
and $\mae[\|\tilde{\deltabf}_{t}\|^2]$ respectively in the following lemma.

\begin{lemma}[Bounding ${\mae[\|\tilde{\alphabf}_{t}\|^2]}$, ${\mae[\|\tilde{\betabf}_{t}\|^2]}$,
and ${\mae[\|\tilde{\deltabf}_{t}\|^2]}$]\label{lemma2}
Consider SegOTA described in Section~\ref{sec:SegOTA} and
joint transmit-receive beamforming described at the beginning of Section~\ref{sec:upper_bound}.
Let $\nu \triangleq \max_{k\in\Kc_{\text{tot}},m\in\Sc_{t},t\in\Tc} \|\tilde{\sbf}^{k,J}_{m,t}\|^2$
and $\tilde{\sigma}^2_{t} \triangleq \sigma^2I_t/2$.
Under Assumption~\ref{assump_bound_diff}, $\mae[\|\tilde{\alphabf}_{t}\|^2]$, $\mae[\|\tilde{\betabf}_{t}\|^2]$,
and $\mae[\|\tilde{\deltabf}_{t}\|^2]$ are respectively upper bounded as
\begin{align}
&\!\! \mae[\|\tilde{\alphabf}_{t}\|^2] \leq \eta^2_tJ^2S_t\phi, \  t\in\Tc, \label{eq_bound2}\\
&\!\! \mae[\|\tilde{\betabf}_{t}\|^2] \leq \eta^2_{t}J^2S_tK^{2}\mu, \  t\in\Tc, \label{eq_bound3}\\
&\!\! \mae[\|\tilde{\deltabf}_{t}\|^2] \leq \nu \sum_{i=1}^{S_t}\frac{\tilde{\sigma}^2_{t}}
{(\sum_{k\in\Kc_{i,t}}\sqrt{p_{k,t}}|\hbf^\textsf{H}_{k,t}\wbf_{i,t}|)^2} \nn\\
&\!\! + \nu
\sum_{i=1}^{S_t}\sum_{j\neq i}K_{j,t}\frac{\sum_{j\neq i}\sum_{q\in\Kc_{j,t}}p_{q,t} |\hbf^\textsf{H}_{q,t}\wbf_{i,t}|^2}
{(\sum_{k\in\Kc_{i,t}}\sqrt{p_{k,t}}|\hbf^\textsf{H}_{k,t}\wbf_{i,t}|)^2}, \  t\in\Tc.  \label{eq_bound4}
\end{align}
\end{lemma}
\IEEEproof
See Appendix~\ref{appA}.
\endIEEEproof

Using the above, we obtain an upper bound on $\mae[\|\thetabf_{T}- \thetabf^\star\|^2]$ in the following proposition.
\begin{proposition}\label{thm:convergence}
For SegOTA described in Section~\ref{sec:SegOTA},  under  Assumptions~\ref{assump_smooth}--\ref{assump_bound_diff} and for
$\eta_t<\frac{1}{L}$, $\forall t\in\Tc$,
the expected model optimality gap after $T$ communication rounds is bounded by
\begin{align}
& \mae[\|\thetabf_{T}- \thetabf^\star\|^2] \nn\\
&\quad\quad\leq \sum_{t=0}^{T-1}\olsi{G}_t\big(H_t(\{\Kc_{i,t}\}, \wbf_t, \pbf_t) + C_t\big) + \Gamma\prod_{t=0}^{T-1}G_t
\label{eq_thm1}
\end{align}
where $\Gamma\triangleq \mae[\| \thetabf_{0} - \thetabf^\star\|^2]$, $G_t \triangleq 4(1-\eta_t\lambda)^{2J}$,
$C_{t} \triangleq 4\eta^2_tJ^2S_t(\phi + K^{2}\mu)$, $\olsi{G}_t \triangleq \prod_{s=t+1}^{T-1}G_s$ with $\olsi{G}_{T-1} = 1$, and
\begin{align}
\!& H_t( \{\Kc_{i,t}\}, \wbf_t, \pbf_t) \triangleq 4\nu\sum_{i=1}^{S_t}\frac{\tilde{\sigma}^2_{t}}
{(\sum_{k\in\Kc_{i,t}}\sqrt{p_{k,t}}|\hbf^\textsf{H}_{k,t}\wbf_{i,t}|)^2}  \nn\\
\!&\qquad + 4\nu
\sum_{i=1}^{S_t}\sum_{j\neq i}K_{j,t}\frac{\sum_{j\neq i}\sum_{q\in\Kc_{j,t}}p_{q,t} |\hbf^\textsf{H}_{q,t}\wbf_{i,t}|^2}
{(\sum_{k\in\Kc_{i,t}}\sqrt{p_{k,t}}|\hbf^\textsf{H}_{k,t}\wbf_{i,t}|)^2}.  \label{eq_H_function}
\end{align}
\end{proposition}
\IEEEproof
Combining \eqref{eq_bound0}--\eqref{eq_bound4}, we have
\begin{align}
& \mae[\|\tilde{\thetabf}_{t+1} - \tilde{\thetabf}^\star\|^2]
\leq G_t\mae[\|\tilde{\thetabf}_{t} - \tilde{\thetabf}^\star\|^2] \nn\\ & \qquad\qquad\qquad\qquad\qquad+H_t( \{\Kc_{i,t}\}, \wbf_t, \pbf_t) + C_t.
\label{eq_full_bound_prop1}
\end{align}
Summing up both sides of \eqref{eq_full_bound_prop1} over $t\in\Tc$ and  rearranging the terms, we have \eqref{eq_thm1}.
\endIEEEproof

\subsection{Beamforming Optimization for SegOTA}\label{sec:joint_group_beam_design}
We replace the objective function in $\Pc_{o}$
with  the upper bound in \eqref{eq_thm1}.
Omitting the constant terms in \eqref{eq_thm1} that do not depend on the optimization variables,
we arrive at an equivalent optimization problem with the objective function $\sum_{t=0}^{T-1}\olsi{G}_t H_t(\{\Kc_{i,t}\}, \wbf_t, \pbf_t)$.
By Proposition~\ref{thm:convergence} and Assumption~\ref{assump_smooth},
we have $\eta_t < \frac{1}{L} \leq \frac{1}{\lambda}$, $\forall t\in\Tc$,
which leads to $G_t > 0$ and further $\olsi{G}_{t} > 0$.
Hence, we separate this optimization problem
into $T$ per-round problems,
each minimizing $H_t(\{\Kc_{i,t}\}, \wbf_t, \pbf_t)$ in communication round $t$.

Next, we apply $\sum_{k\in\Kc_{i,t}}p_{k,t}|\hbf^\textsf{H}_{k,t}\wbf_{i,t}|^2 \le (\sum_{k\in\Kc_{i,t}}\sqrt{p_{k,t}}|\hbf^\textsf{H}_{k,t}\wbf_{i,t}|)^2$ to the denominators in \eqref{eq_H_function}
to further upper bound $H_t(\{\Kc_{i,t}\}, \wbf_t, \pbf_t)$.
Then, we arrive at the following per-round online optimization problem:
\begin{align}
\Pc_{1,t}: \!\!\!\!\!\!\min_{\{\Kc_{i,t}\}^{S_t}_{i=1}, \wbf_t, \pbf_t}  &
\sum_{i=1}^{S_t}\frac{Z_{i,t}\sum_{j\neq i}\sum_{q\in\Kc_{j,t}}p_{q,t} |\fbf^\textsf{H}_{q,t}\wbf_{i,t}|^2 + 1}
{\sum_{k\in\Kc_{i,t}}p_{k,t}|\fbf^\textsf{H}_{k,t}\wbf_{i,t}|^2} \nn\\
\text{s.t.} \ \ & \Kc_{i,t}\bigcap\Kc_{j,t} = \varnothing,\quad i\neq j,\,\, i,j\in\Sc_t,   \nn\\
& \bigcup\limits_{i\in\Sc_t}\Kc_{i,t} = \Kc_{\text{tot}},\nn\\
& p_{k,t} \le I_tP_k, \quad k\in \Kc_{\text{tot}},\nn\\
& \|\wbf_{i,t}\|^2 = 1, \quad i\in \Sc_t \nn
\end{align}
where $Z_{i,t} \triangleq \sum_{j\neq i}K_{j,t}$ and $\fbf_{k,t} \triangleq \hbf_{k,t}/\tilde{\sigma}_{t}$.
Note that the objective function represents a sum of the inverse of received signal-to-interference-and-noise ratio (SINR) corresponding to received aggregated segment from each device group.

The above is a mixed-integer programming problem. Furthermore, the objective function is nonconvex
w.r.t. the power vector $\pbf_t$ and the beamforming vector $\wbf_t$,
which is challenging to solve.
We propose a device grouping scheme based on
spherical {\it k}-means to first obtain $\{\Kc_{i,t}\}$. Based on this, we
then optimize the joint transmit-receive beamforming $(\wbf_t,\pbf_t)$
in $\Pc_{1,t}$.

\subsubsection{Device grouping via spherical {\it k}-means}
\label{sec:kmeans_device_grouping}
Since receive beamforming $\wbf_{i,t}$ is applied to the devices of the same group $i$,
more spatially correlated device channels can lead to
higher received beamforming gain for this group \cite{Kiani&etal:TComm22}.
For this purpose, we propose a device grouping scheme that uses the clustering idea to
find the spatially correlated device groups.
The scheme is based on the spherical {\it k}-means framework \cite{Dhillon&Modha:ML01},
which is a variant of the standard {\it k}-means that captures the cosine similarity
among data points to form clusters.
In particular, we first define the feature space for the purpose of device grouping.
Let $\Xc_t$ denote the feature space spanned by the device uplink channels in round $t$, given by
\begin{align}
\Xc_t = \left\{\xbf_{k,t}: \xbf_{k,t} \triangleq \frac{\hbf_{k,t}}{\|\hbf_{k,t}\|}e^{-j\angle{h_{1k,t}}},\, \forall k\in\Kc_{\text{tot}}\right\} \nn
\end{align}
where $\angle{h_{1k,t}}$ denotes the phase of the first element in $\hbf_{k,t}$.
Each data point $\xbf_{k,t}$ in $\Xc_t$ is phase-adjusted such that
its first element is phase-aligned to $0$ degree.
This is to ensure that during the iterative updating process,
all $\xbf_{k,t}$'s are phase-aligned to sum up properly in computing the centroid.
Let $\cbf_{r,t}$ denote the centroid of cluster $r=1,\ldots,S_t$ in $\Xc_t$ with $\|\cbf_{r,t}\|=1$.
We consider the following metric to measure distance from
each data point $\xbf_{k,t}$ in $\Xc_t$ to a centroid $\cbf_{r,t}$:
\begin{align}
\delta(\xbf, \cbf) = |\xbf_{k,t}^\textsf{H} \cbf_{r,t}|, \ \forall \xbf_{k,t} \in
\Xc_t   \label{eq_kmeans_clustering}
\end{align}
where $\delta(\xbf, \cbf)\in [0, 1]$. This distance metric measures the correlation level between the channel vector and the centroid,
with $1$ being fully correlated
and $0$ being orthogonal.
A data point $\xbf_{k,t}$ will be included in cluster $\cbf_{r,t}$ where it has
the largest $\delta(\xbf, \cbf)$ among all clusters.
Specifically, denote the set of $\xbf_{k,t}$'s in cluster $\cbf_{r,t}$ by
\begin{align}
\Yc_{r,t} = \{ \xbf_{k,t}\in\Xc_t: \delta(\xbf_{k,t}, \cbf_{r,t}) > \delta(\xbf_{k,t}, \cbf_{r'\!\!,t}), \ r'\neq r\}.  \nn
\end{align}
Given $\Xc_t$ and $\delta(\xbf, \cbf)$, we then apply the spherical {\it k}-means method \cite{Dhillon&Modha:ML01} to form $S_t$
centroid points and clusters in $\Xc_t$ and iteratively update the centroid points $\cbf_{r,t}$'s.
The centroid update $\cbf^{(l+1)}_{r,t}$ at iteration $l$ is then given by
\begin{align}
\cbf^{(l+1)}_{r,t} =
\frac{\sum_{\xbf_{k,t}\in\Yc_{r,t}} \xbf_{k,t}}{  |\Yc_{r,t}|}; \;\; \cbf^{(l+1)}_{r,t} \leftarrow
\frac{\cbf^{(l+1)}_{r,t}}{\|\cbf^{(l+1)}_{r,t}\|}. \label{eq:ms_centroid_update_truncat}
\end{align}
The above procedure is repeated until convergence to obtain a device grouping solution $\{\Kc_{i,t}\}^{S_t}_{i=1}$.

\subsubsection{Joint transmit-receive beamforming}\label{sec:uplink_joint_BF}
Given $\{\Kc_{i,t}\}^{S_t}_{i=1}$,
we now optimize uplink beamforming $(\wbf_t,\pbf_t)$
in $\Pc_{1,t}$.
Since the problem is nonconvex,
we propose to
alternatingly optimize receive beamforming $\wbf_t$
and the device transmit powers in $\pbf_t$ via the
block coordinate descent (BCD) method \cite{Bertsekas:2016nonlinear}.
The two subproblems are given below:

\emph{i) Updating $\wbf_t$:}
Given $\pbf_t$, $\Pc_{1,t}$ can be equivalently decomposed into $S_t$ subproblems,
one for each beamformer  $\wbf_{i,t}$ for each device group $i$ as
\begin{align}
\Pc^{\text{wsub1}}_{1,t,i}: \ \ & \min_{\wbf_{i,t}}
 \frac{Z_{i,t}\sum_{j\neq i}\sum_{q\in\Kc_{j,t}}p_{q,t} |\fbf^\textsf{H}_{q,t}\wbf_{i,t}|^2 \!+\! 1}
{\sum_{k\in\Kc_{i,t}}p_{k,t}|\fbf^\textsf{H}_{k,t}\wbf_{i,t}|^2}   \nn\\
&\ \text{s.t.} \ \ \|\wbf_{i,t}\|^2 = 1.  \nn
\end{align}
After expanding the quadratic terms in $\Pc^{\text{wsub1}}_{1,t,i}$,
$\wbf_{i,t}$ can be moved outside of the summation at both the numerator and denominator, given by
\begin{align}
\Pc^{\text{wsub2}}_{1,t,i}: \ \ & \min_{\wbf_{i,t}}
 \frac{\wbf^{\textsf{H}}_{i,t} \left( Z_{i,t}\sum_{j\neq i}\sum_{q\in\Kc_{j,t}}p_{q,t}\fbf_{q,t}\fbf^{\textsf{H}}_{q,t}  + \Ibf \right)
 \wbf_{i,t}}{\wbf^{\textsf{H}}_{i,t} \left( \sum_{k\in\Kc_{i,t}}p_{k,t}\fbf_{k,t}\fbf^{\textsf{H}}_{k,t} \right)
 \wbf_{i,t}}   \nn\\
&\ \text{s.t.} \ \ \|\wbf_{i,t}\|^2 = 1,  \nn
\end{align}
which is a generalized eigenvalue problem.
Specifically,
let $\Abf_{i,t} \triangleq Z_{i,t}\sum_{j\neq i}\sum_{q\in\Kc_{j,t}}p_{q,t}\fbf_{q,t}\fbf^{\textsf{H}}_{q,t}  + \Ibf$
and $\Bbf_{i,t} \triangleq \sum_{k\in\Kc_{i,t}}p_{k,t}\fbf_{k,t}\fbf^{\textsf{H}}_{k,t}$.
We re-express problem $\Pc^{\text{wsub2}}_{1,t,i}$ as
\begin{align}
\Pc^{\text{wsub}3}_{1,t,i}: \,\, \min_{\wbf_{i,t}} \,\,  \frac{\wbf^{\textsf{H}}_{i,t}\Abf_{i,t}\wbf_{i,t}}
{\wbf^{\textsf{H}}_{i,t}\Bbf_{i,t}\wbf_{i,t}} \quad \text{s.t.} \ \  \|\wbf_{i,t}\|^2 = 1. \nn
\end{align}
The optimal solution $\wbf_{i,t}$ of $\Pc^{\text{wsub}3}_{1,t,i}$ is the generalized eigenvector corresponding to
the smallest generalized eigenvalue in the generalized eigenvalue problem of
$\Abf_{i,t}\Phibf_{i,t} = \Bbf_{i,t}\Phibf_{i,t}\Lambdabf_{i,t}$, where $\Phibf_{i,t}$ is the eigenvector matrix
and $\Lambdabf_{i,t}$ a diagonal matrix with its diagonal elements being the eigenvalues.
Let $\Phibf^{\text{A}}_{i,t}$ and $\Phibf^{\text{B}}_{i,t}$ denote the eigenvector matrices of $\Abf_{i,t}$ and $\Bbf_{i,t}$,
respectively.
Let $\Lambdabf^{\text{B}}_{i,t}$ be a diagonal matrix with its diagonal elements being the eigenvalues of $\Bbf_{i,t}$.
Then, the generalized eigenvector matrix is given by
\begin{align}
 \Phibf_{i,t} = \Phibf^{\text{B}}_{i,t}(\Lambdabf^{\text{B}}_{i,t})^{-1/2}\Phibf^{\text{A}}_{i,t}.  \label{eq_uplink_BF_solution}
\end{align}
The optimal solution $\wbf_{i,t}$ is a column vector in $\Phibf_{i,t}$ corresponding to the smallest generalized eigenvalue.

\emph{ii) Updating $\pbf_t$:}
Let $\gbf_{ij,t}$ be the vector containing $\{g_{iq,t}\triangleq |\fbf^\textsf{H}_{q,t}\wbf_{i,t}|^2$,  $q\in\Kc_{j,t}\}$ from group $j$ after applying receive beamformer
 $\wbf_{i,t}$.
 Given $\wbf_t$, we can rewrite $\Pc_{1,t}$ as
\begin{align}
\Pc^{\text{psub}1}_{1,t}: \,\,\min_{\pbf_t} \,\, &
\sum_{i=1}^{S_t}\frac{Z_{i,t}\sum_{j\neq i}\gbf^\textsf{T}_{ij,t}\pbf_{j,t} +1}{\gbf^\textsf{T}_{ii,t}\pbf_{i,t}}\nn\\
\text{s.t.} \ \ & p_{k,t} \le I_tP_k, \quad k\in \Kc_{\text{tot}}. \nn
\end{align}
To efficiently compute $\pbf_t$, we adopt BCD
to update $\pbf_{1,t},\ldots,\pbf_{S_t,t}$
alternatingly, one for each group $i$.
Specifically, given $\pbf_{j,t}$, $\forall j\in\Sc_t, j\neq i$,
the optimization of $\pbf_{i,t}$ for group $i$ is given by
\begin{align}
\Pc^{\text{psub}2}_{1,t,i}: \,\,\min_{\pbf_{i,t}} \,\, &
\frac{Z_{i,t}\sum_{j\neq i}\gbf^\textsf{T}_{ij,t}\pbf_{j,t} +1}{\gbf^\textsf{T}_{ii,t}\pbf_{i,t}} + \sum_{j\neq i} \frac{Z_{j,t}\gbf^\textsf{T}_{ji,t}}{\gbf^\textsf{T}_{jj,t}\pbf_{j,t}}\pbf_{i,t}\nn\\
\text{s.t.} \ \ & p_{k,t} \le I_tP_k, \quad k\in \Kc_{i,t}. \nn
\end{align}
Problem $\Pc^{\text{psub}2}_{1,t,i}$ is convex w.r.t. $\pbf_{i,t}$,
for which the optimal $\pbf_{i,t}$ can be obtained in closed-form.
Specifically, let
\begin{align}
\beta^{\text{min}}_{i,t} \triangleq \min_{k\in\Kc_{i,t}}\bigg(\frac{Z_{i,t}\sum_{j\neq i}\gbf^\textsf{T}_{ij,t}\pbf_{j,t} +1}{ \sum_{j\neq i}\frac{Z_{j,t}g_{jk,t}} {\gbf^\textsf{T}_{jj,t}\pbf_{j,t}}}g_{ik,t}\bigg)^{1/2},\nn
\end{align}
and let $k'\in \Kc_{i,t}$ be the corresponding device index that achieves $\beta^{\text{min}}_{i,t}$.
Let $\olsi{\Pbf}_{i}$ be the vector containing the maximum power of devices in group $i$
$\{P_k, k\in\Kc_{i,t}\}$.
Then, each device transmit power $p_{k,t}, k\in\Kc_{i,t}$ in the optimal $\pbf_{i,t}$ is given by
\begin{align}
p_{k,t} =
\begin{cases}
\displaystyle I_tP_k     &\!\!\!\!\!\!\!\!\!\!\!\!\!\!\!\!\!\!\!\!\!\!\! \text{for~} k\in\Kc_{i,t}, k\neq k' \\
\displaystyle I_tP_{k'}  - \frac{\displaystyle \left[I_t\gbf^\textsf{T}_{ii,t}\olsi{\Pbf}_{i}-\beta^{\text{min}}_{i,t}\right]^+}{\displaystyle g_{ik',t}}     & \text{for~} k=k' \nn
\end{cases}
\end{align}
where $[\beta]^+ =\max\{\beta,0\}$.
All $\pbf_{i,t}$'s are updated alternatingly using the above solution.

Since subproblems $\Pc^{\text{wsub}3}_{1,t,i}$ and
$\Pc^{\text{psub}2}_{1,t,i}$ for each $i$ are solved optimally, and our optimization objective is lower bounded by zero, by the monotone convergence theorem of BCD \cite{Bertsekas:2016nonlinear}, our proposed algorithm for computing $(\wbf_t,\pbf_t)$ is guaranteed to converge
to a stationary point.

\section{Simulation Results}
\label{sec:simulations}

\subsubsection{Simulation Setup}
We evaluate our proposed SegOTA for FL  on image classification over a simulated wireless network under typical wireless specifications
with system  bandwidth  $1$~MHz, carrier frequency $2$~GHz,
and per-device maximum transmit power at each device $P_k = 23~\text{dBm}$.
We generate channels as $\hbf_{k,t} =
\sqrt{G_{k}}\bar{\hbf}_{k,t}$, where
$\bar{\hbf}_{k,t}\sim\mathcal{CN}({\bf{0}},{\bf{I}})$ and the path gain
$G_{k} [\text{dB}] = -136.3-35\log_{10}d_k - \psi_k$,
with  BS-device distance $d_k$ being uniformly distributed within $(0.02, 0.5)~\text{km}$ and shadowing variable
$\psi_k \sim \Nc(0, \sigma_\psi^2)$
with $\sigma_\psi = 8~\text{dB}$.
We set the receiver noise power $\sigma^2 = -79~\text{dBm}$,
which accounts for both thermal noise and inter-cell interference.

We use the MNIST dataset for model training and testing.
It consists of $6\times 10^4$ training samples and $1\times 10^4$ test samples.
We trains a convolutional neural network with an $8\times3\times3$ ReLU convolutional layer,
a $2\times2$ max pooling layer, a ReLU fully-connected layer with $20$ units, and a softmax output layer,
with $D=2.735\times 10^4$ model parameters.
We use the $ 10^4$ test samples to measure the test accuracy of the global model update $\thetabf_t$ at each round $t$.
The training samples are randomly and evenly distributed  across $K$ devices, with local dataset size $A_{k} = \frac{6\times 10^4}{K}$ at device $k$.
For local training using SGD, we set $J=100$, mini-batch size $|\Bc^{\tau}_{k,t}|={600}/{K}, \forall t,\tau,k$, and learning rate $\eta_t=0.1$, $\forall t$.
All results are obtained by averaging over $20$ channel realizations.

\begin{figure}[t]
\centering
\includegraphics[width=0.88\columnwidth]{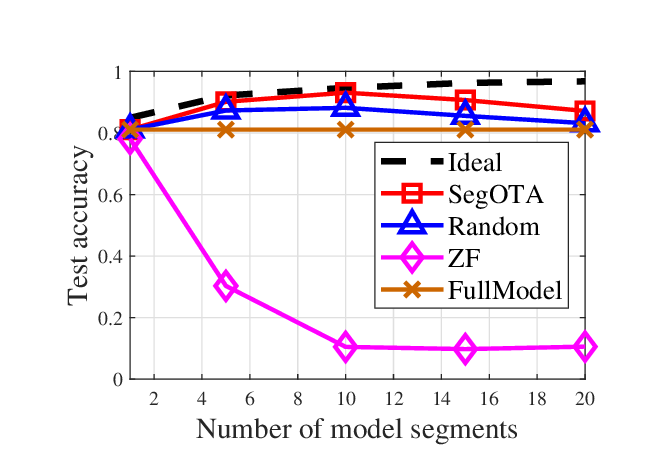}
\vspace{-1.3em}\caption{Test accuracy vs. number of model segments $S_t$ ($(N,K) = (32,50)$).}
\label{Fig2:Acc_St_N32K50}\vspace*{-1.3em}
\end{figure}

\begin{figure}[t]
\centering
\includegraphics[width=0.88\columnwidth]{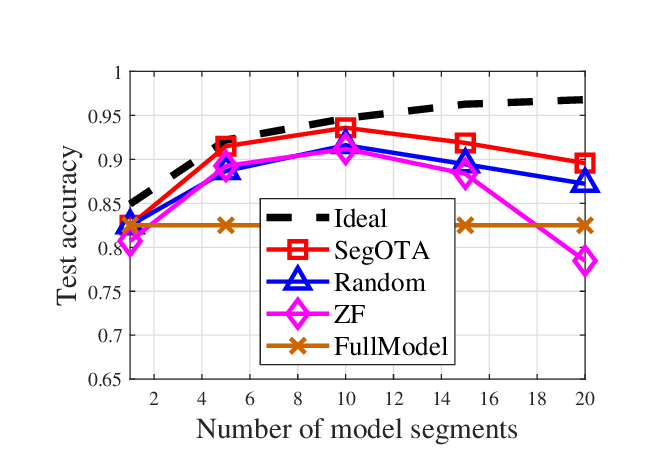}
\vspace{-1.3em}\caption{Test accuracy vs. number of model segments $S_t$ ($(N,K) = (64,50)$).}
\label{Fig3:Acc_St_N64K50}\vspace*{-1.5em}
\end{figure}

\begin{figure}[t]
\centering
\includegraphics[width=0.88\columnwidth]{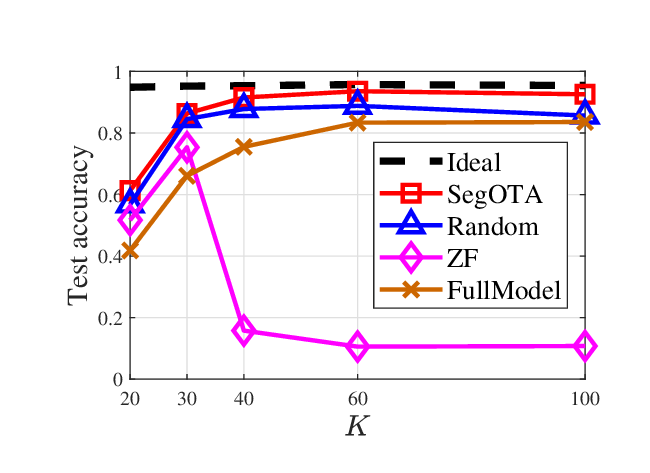}
\vspace{-1.3em}\caption{Test accuracy vs. number of devices $K$ for $N=32$ and $S_t=10$.}
\label{Fig4:Acc_K_N32St10}\vspace*{-1.5em}
\end{figure}

\subsubsection{Performance Comparison}
 For comparison, we consider the following five schemes:
i) \textbf{Ideal}: SegOTA via  \eqref{eq_global_update_0} with noise-interference-free uplink and perfect recovery of model parameters, which serves as a performance upper bound for all schemes.
ii) \textbf{SegOTA}: our proposed method.
iii) \textbf{Random}: the same as the proposed SegOTA, except that devices are randomly partitioned
into equal-sized groups instead of our grouping method proposed in
Section~\ref{sec:kmeans_device_grouping}.
iv) \textbf{ZF}: the same as the proposed SegOTA and device grouping in
Section~\ref{sec:kmeans_device_grouping},
except that we apply zero-forcing receive beamforming under maximum device transmit powers \cite{Sadeghi&etal:2018TWC}.
v) \textbf{FullModel}: traditional full-model OTA approach, which is equivalent to  SegOTA with $S_t = 1, \forall t$.

We study the tradeoff between communication efficiency in the uplink model transmission
and test accuracy of the global model updates.
Figs.~\ref{Fig2:Acc_St_N32K50}--\ref{Fig4:Acc_K_N32St10} compare the test accuracy performance of the five methods
using a total of $2.736\times 10^4$ channel uses per device for uplink model transmission.
Each device group is randomly assigned a unique segment, as mentioned at the beginning of Section~\ref{sec:SegOTA}.
Fig.~\ref{Fig2:Acc_St_N32K50} and \ref{Fig3:Acc_St_N64K50} show the test accuracy vs.\   $S_t$ model segments for
the overloaded setup $(N,K)=(32,50)$ and the underloaded setup $(N,K)=(64,50)$, respectively.  SegOTA outperforms both Random and ZF for all $S_t>1$.
Furthermore, SegOTA nearly attains the performance of Ideal for
$S_t\leq 10$.
Fig.~\ref{Fig4:Acc_K_N32St10} shows the test accuracy vs.\ $K$ devices for $N=32$ and $S_t =10$.
Again, we see that SegOTA nearly attains the optimal performance under Ideal and outperforms the other alternatives.
We also observe that ZF can perform worse than FullModel, especially when $S_t$ or $K$ is large. This shows the effectiveness of our proposed beamforming algorithm for SegOTA targeting at FL training performance, while conventional ZF for interference cancellation may not be effective.

\vspace*{-.5em}
\section{Conclusion}\label{sec:conclusion}
This paper proposes a segmented transmission approach SegOTA to reduce the latency in uplink OTA aggregation for wireless FL.
Under SegOTA, devices are divided into groups, where each group
is assigned a segment of the model for OTA aggregation, and all segments are sent via uplink simultaneously. Based on the segment global updating equation, we derive an upper bound on the model training optimality gap to formulate a joint transmit-receive beamforming problem along with device grouping. We propose a device grouping scheme based on their spatial channel correlations via spherical {\it k}-means  and an iterative uplink beamforming algorithm with fast closed-form updates.
Simulation results show the proposed SegOTA with proposed device grouping and uplink beamforming outperforms the traditional full-model OTA approach and other alternatives.

\vspace*{-.5em}
\appendices
\section{Proof of Lemma~\ref{lemma2}}\label{appA}
\IEEEproof
For bounding $\mae[\|\tilde{\alphabf}_{t}\|^2]$, we have
\begin{align}
&\mae[\|\tilde{\alphabf}_{t}\|^2]  \stackrel{(a)}{=} \eta^2_{t}\sum^{S_t}_{m=1}\mae\Bigg[\Bigg\|\sum^{J-1}_{\tau=0}\nabla \tilde{F}^m(\vbf^{\tau}_{t})   \nn\\
&\qquad\qquad\quad - \sum_{k\in\Kc_{\hat{i}(m,t),t}}\rho_{k,t}\sum_{\tau=0}^{J-1} \nabla \tilde{F}_{k}^{m}(\thetabf^{\tau}_{k,t})\Bigg\|^2\Bigg]  \nn\\
& \leq \eta^2_{t}\sum^{S_t}_{m=1}\mae\Bigg[\Bigg(\sum^{J-1}_{\tau=0}\bigg\| \nabla \tilde{F}^m(\vbf^{\tau}_{t}) \nn\\
&\qquad\qquad\quad - \sum_{k\in\Kc_{\hat{i}(m,t),t}}\rho_{k,t} \nabla \tilde{F}_{k}^{m}(\thetabf^{\tau}_{k,t})\bigg\|
\Bigg)^2\Bigg]  \nn\\
& \stackrel{(b)}{\leq} \eta^2_{t}J\sum^{S_t}_{m=1}\sum^{J-1}_{\tau=0}
\mae\bigg[\bigg\|\nabla \tilde{F}^m(\vbf^{\tau}_{t}) \nn\\
&\qquad\qquad\quad - \sum_{k\in\Kc_{\hat{i}(m,t),t}}\rho_{k,t} \nabla \tilde{F}_{k}^{m}(\thetabf^{\tau}_{k,t})\bigg\|^2\bigg]\nn\\
& \stackrel{(c)}{\leq}  \eta^2_tJ^2S_t\phi. \nn
\end{align}
where $(a)$ uses the expression of $\tilde{\alphabf}_{t}$ in \eqref{eq_frame_update_derive},
$(b)$ is based on $( \sum^{G}_{i=1}x_{i} )^2 \leq G\sum^{G}_{i=1}x^2_{i}, \forall x_{i}\in\mathbb{R}$, for some $G\in\mathbb{N}^{+}$,
and $(c)$ follows Assumption~\ref{assump_bound_diff}.
Thus, we have \eqref{eq_bound2}.

For bounding $\mae[\|\tilde{\betabf}_{t}\|^2]$, we have
\begin{align}
&\mae[\|\tilde{\betabf}_{t}\|^2]   \stackrel{(a)}{=} \eta^2_{t}\sum^{S_t}_{m=1}\mae\Bigg[\Bigg\|
\sum_{k\in\Kc_{\hat{i}(m,t),t}}\rho_{k,t}\sum_{\tau=0}^{J-1}\Big(\nabla \tilde{F}^{m}_{k}(\thetabf^{\tau}_{k,t}) \nn\\
&\qquad\qquad\quad - \nabla \tilde{F}^{m}_{k}(\thetabf^{\tau}_{k,t}; \Bc^{\tau}_{k,t}) \Big)\Bigg\|^2\Bigg] \nn\\
& \leq   \eta^2_{t}\sum^{S_t}_{m=1}\mae\Bigg[\Bigg(
\sum_{k\in\Kc_{\hat{i}(m,t),t}}\sum_{\tau=0}^{J-1}|\rho_{k,t}|\Big\|\nabla \tilde{F}^{m}_{k}(\thetabf^{\tau}_{k,t}) \nn\\
&\qquad\qquad\quad - \nabla \tilde{F}^{m}_{k}(\thetabf^{\tau}_{k,t}; \Bc^{\tau}_{k,t}) \Big\|\Bigg)^2\Bigg] \nn\\
& \stackrel{(b)}{\leq} \eta^2_{t}\sum^{S_t}_{m=1}  \sum_{k\in\Kc_{\hat{i}(m,t),t}}\sum_{\tau=0}^{J-1}
JK_{\hat{i}(m,t),t}
\mae\Big[\Big\|\nabla \tilde{F}^{m}_{k}(\thetabf^{\tau}_{k,t}) \nn\\
&\qquad\qquad\quad - \nabla \tilde{F}^{m}_{k}(\thetabf^{\tau}_{k,t}; \Bc^{\tau}_{k,t}) \Big\|^2\Big]  \nn\\
& \stackrel{(c)}{\leq} \eta^2_{t}J^2S_tK^{2}\mu   \nn
\end{align}
where $(a)$ uses the expression of $\tilde{\deltabf}_{t}$ in \eqref{eq_frame_update_derive},
$(b)$ is based on $( \sum^{G}_{i=1}x_{i} )^2 \leq G\sum^{G}_{i=1}x^2_{i}, \forall x_{i}\in\mathbb{R}$, for some $G\in\mathbb{N}^{+}$,
and $(c)$ follows Assumption~\ref{assump_bound_diff}.
Thus, we have \eqref{eq_bound3}.

For bounding $\mae[\|\tilde{\deltabf}_{t}\|^2]$, we have
\begin{align}
&\mae[\|\tilde{\deltabf}_{t}\|^2]  \stackrel{(a)}{=} \sum^{S_t}_{m=1}\mae\Bigg[\Bigg\|\tilde{\nbf}_{m,t} + \frac{1}{\alpha^{\text{s}}_{m,t}} \nn\\
& \cdot \sum_{j\neq \hat{i}(m,t)}\sum_{q\in\Kc_{j,t}}\!\!\! \frac{\hbf^{\textsf{H}}_{q,t}\wbf_{j,t}\wbf_{\hat{i}(m,t),t}^\textsf{H}
\hbf_{q,t}}{|\hbf^{\textsf{H}}_{q,t}\wbf_{j,t}|}\cdot \frac{\sqrt{p_{q,t}}\tilde{\sbf}^{q,J}_{\hat{m}(j, t),t}}{\|\tilde{\sbf}^{q,J}_{\hat{m}(j, t),t}\|}\Bigg\|^2\Bigg] \nn\\
&\leq \sum^{S_t}_{m=1}\mae\Bigg[\Bigg(\|\tilde{\nbf}_{m,t}\|+ \frac{1}{\alpha^{\text{s}}_{m,t}} \nn\\
& \cdot \sum_{j\neq \hat{i}(m,t)}\sum_{q\in\Kc_{j,t}}\! \Bigg\|\frac{\hbf^{\textsf{H}}_{q,t}\wbf_{j,t}\wbf_{\hat{i}(m,t),t}^\textsf{H}
\hbf_{q,t}}{|\hbf^{\textsf{H}}_{q,t}\wbf_{j,t}|}\cdot \frac{\sqrt{p_{q,t}}\tilde{\sbf}^{q,J}_{\hat{m}(j, t),t}}{\|\tilde{\sbf}^{q,J}_{\hat{m}(j, t),t}\|}\Bigg\|\Bigg)^2\Bigg] \nn\\
&\stackrel{(b)}{\leq} \nu \sum_{i=1}^{S_t}\frac{\tilde{\sigma}^2_{t}}
{(\sum_{k\in\Kc_{i,t}}\sqrt{p_{k,t}}|\hbf^\textsf{H}_{k,t}\wbf_{i,t}|)^2} \nn\\
& + \nu
\sum_{i=1}^{S_t}\sum_{j\neq i}K_{j,t}\frac{\sum_{j\neq i}\sum_{q\in\Kc_{j,t}}p_{q,t} |\hbf^\textsf{H}_{q,t}\wbf_{i,t}|^2}
{(\sum_{k\in\Kc_{i,t}}\sqrt{p_{k,t}}|\hbf^\textsf{H}_{k,t}\wbf_{i,t}|)^2} \nn
\end{align}
where $(a)$ uses the expression of $\tilde{\deltabf}_{t}$
in \eqref{eq_frame_update_derive},
and $(b)$ is based on
$( \sum^{G}_{i=1}x_{i} )^2 \leq G\sum^{G}_{i=1}x^2_{i}, \forall x_{i}\in\mathbb{R}$, for some $G\in\mathbb{N}^{+}$.
Thus, we have \eqref{eq_bound4}.
\endIEEEproof

\bibliographystyle{IEEEbib}
\bibliography{Refs}

\end{document}